\documentstyle[aps,prl,twocolumn,graphicx,floats,epsfig]{revtex}
\begin{document}
\draft
\onecolumn 

\noindent

\title{Centrality Dependence of Neutral Pion Production in 158$\cdot$A~GeV 
$^{208}$Pb\/+\/$^{208}$Pb Collisions}

\author{M.~M.~Aggarwal,$^{1}$
A.~Agnihotri,$^{2}$
Z.~Ahammed,$^{3}$
A.~L.~S.~Angelis,$^{4}$ 
V.~Antonenko,$^{5}$ 
V.~Arefiev,$^{6}$
V.~Astakhov,$^{6}$
V.~Avdeitchikov,$^{6}$
T.~C.~Awes,$^{7}$
P.~V.~K.~S.~Baba,$^{8}$
% ...
}
\author{WA98 Collaboration}

\address{$^{1}$~University of Panjab, Chandigarh 160014, India}
\address{$^{2}$~University of Rajasthan, Jaipur 302004, Rajasthan,
  India}
\address{$^{3}$~Variable Energy Cyclotron Centre,  Calcutta 700 064,
  India}
\address{$^{4}$~University of Geneva, CH-1211 Geneva 4,Switzerland}
\address{$^{5}$~RRC (Kurchatov), RU-123182 Moscow, Russia}
\address{$^{6}$~Joint Institute for Nuclear Research, RU-141980 Dubna,
  Russia}
\address{$^{7}$~Oak Ridge National Laboratory, Oak Ridge, Tennessee
  37831-6372, USA}
\address{$^{8}$~University of Jammu, Jammu 180001, India}
% ...

\date{\today}
\maketitle
\begin{abstract}
The production of neutral pions in 158$\cdot$A~GeV 
$^{208}$Pb\/+\/$^{208}$Pb  collisions
has been studied in the WA98 experiment at the CERN SPS.
Transverse momentum spectra are studied for the range
 $0.3 \, {\mathrm{GeV}}/c \le m_{T} - m_{0} \le 4.0 \, {\mathrm{GeV}}/c$.
The results for central collisions are compared to various models.
The centrality dependence of the neutral pion spectral shape and 
yield is investigated.
An invariance of the spectral shape and a simple
scaling of the yield with the number of participating nucleons is observed 
for centralities with greater than about 30 participating nucleons 
which is most naturally explained by assuming an equilibrated system.

\end{abstract}
\pacs{25.75.Dw}
\twocolumn

Ultra-relativistic heavy-ion collisions produce dense matter which 
is expected to be in the form of a deconfined phase of quarks and gluons
at sufficiently high energy densities.
The transverse momentum spectra of produced pions can provide 
information on both the initial and final state properties of the 
hot hadronic matter. The  high $p_{T}$ pion production is expected to be 
dominated by hard scattering of the partons. In pA collisions the 
high $p_{T}$ region is known to be enhanced (Cronin effect
\cite{cronin}) due to initial state
scattering of the incident partons leading to a broadening of their
incoming $p_{T}$. In AA collisions, many of the scattered partons
must traverse the excited matter to escape and therefore may undergo
additional rescatterings and energy loss \cite{wang92}. In the case
of significant parton rescattering, the parton distributions may
approach thermal distributions with a temperature reflecting the 
initial state of the excited matter. The intermediate $p_{T}$ 
region of the pion spectrum might then reflect this initial
temperature while the low $p_{T}$ region would dominantly reflect the
freeze-out temperature. Indeed, one of the earliest signatures of
QGP formation, proposed by Van Hove \cite{hove82},
was the observation of a saturation of the average transverse momentum
with increasing energy (or entropy) density for systems excited 
just above the critical energy density. With increasing energy
density, the initial temperature would not rise above the critical 
temperature until all of the latent heat of the QQP phase transition
had been extracted.

For these reasons it is of interest to study the centrality dependence
of the pion production. It is generally believed that the initial 
energy density increases with increasing centrality, due to the many
overlapping interactions. Also, the volume of the excited matter
increases with centrality, as well as the amount of rescattering.
Since rescattering is the feature which distinguishes
AA collisions non-trivially from pp collisions, and since significant
rescattering is a prerequisite for thermalization, it is 
imperative to demonstrate an understanding of the centrality
dependence of the AA results in order to understand
the effects of rescattering. While those effects may
be minor on extensive observables, like the particle multiplicity or 
transverse energy, they should be most evident on the momentum 
distribution of the produced particles.  Recently it has been argued that a
parton cascade description could successfully describe many of the
features of central Pb+Pb collisions at SPS energies \cite{GEIGER97}.
Surprisingly, low momentum transfer soft parton collisions were found to
have little influence on the final observables. 
Similarly, recent perturbative QCD
calculations 
were able to reproduce the preliminary 
WA98 neutral pion result for central collisions 
\cite{Peitzmann:1996:qm96,Peitzmann:1997:qm97} 
without need for the effects of parton
energy loss or rescattering \cite{wang:1998:qcd}.
In this letter we present neutral pion spectra for 158$\cdot$A~GeV 
$^{208}$Pb\/+\/$^{208}$Pb  collisions and investigate in detail the centrality 
dependence of the spectral shape and yield.

The CERN experiment WA98 \cite{Peitzmann:1996:qm96,misc:wa98:proposal:91} 
consists
of large acceptance photon and hadron spectrometers together with several 
other large acceptance devices which allow to measure various global
variables on an event-by-event basis. 
The results presented here were obtained from an analysis of the
data taken with Pb beams in 1995 and 1996.
The minimum bias reactions ($\sigma_{min.bias} \approx 6300 \, \mathrm{mb}$) 
are divided into eight centrality classes using the transverse 
energy $E_{T}$ measured in the MIRAC calorimeter. 
In total, $\approx 9.6 \cdot 10^{6}$ reactions have been analyzed.

Neutral pions are reconstructed via their $\gamma\gamma$ decay branch 
using the WA98 lead-glass photon detector,
LEDA, which consisted of 10,080 individual modules with 
photomultiplier readout. The detector was located at a distance of 
21.5~m from the target and covered the pseudorapidity interval 
$2.35 < \eta < 2.95$.

The general analysis procedure is similar to that used in the 
WA80 experiment and described in \cite{wa80:pi0:98}. Hits in the 
lead-glass detector are combined in pairs to provide distributions of pair 
mass vs. pair transverse momentum (or transverse mass) 
for all possible combinations. 
Subtraction of the combinatorial background is performed using 
mixed event distributions. The resulting momentum 
distributions are corrected for geometrical acceptance and  
reconstruction efficiency. The efficiency depends on the particle 
occupancy in the detector and therefore has been calculated 
independently for each centrality bin. The systematic error of the 
pion yields is mainly due to errors in the reconstruction 
efficiency for central collisions and to corrections for non-target
interactions for peripheral collisions. 
The systematic error on the absolute yield is $\approx 10 \%$ and
increases sharply below $p_{T}  = 0.4 \, {\mathrm{GeV}}/c$.
An additional systematic error originates 
from the uncertainty of the momentum scale of 1~\%. The influence of 
this rises slowly for higher $p_{T}$ and leads to an error of 15~\% 
at $p_{T} = 4 \, {\mathrm{GeV}}/c$. A detailed discussion 
of the analysis procedure and the error contributions will be given in 
a forthcoming publication.

\begin{figure}[bt]
        \centerline{\includegraphics{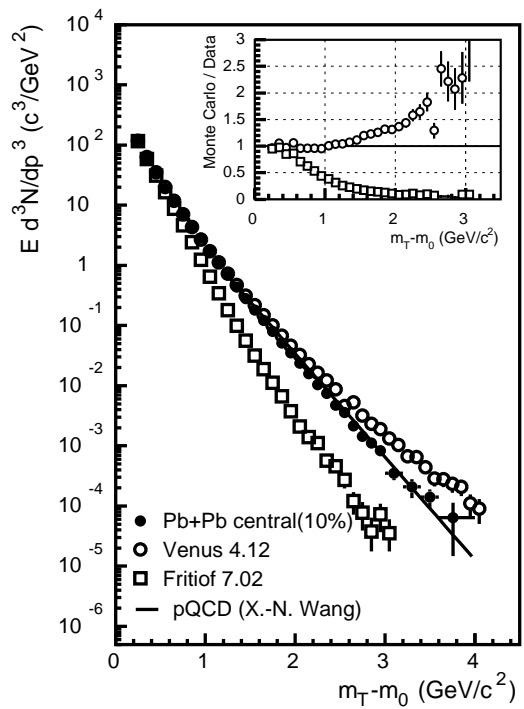}}
        \caption{Transverse mass spectra of neutral pions in central 
        collisions of 158 $A$GeV Pb+Pb. Invariant 
        yields per event are compared to calculations using the FRITIOF 
        7.02 \protect\cite{ANDERSSON93} and 
        VENUS 4.12 \protect\cite{WERNER93a} Monte Carlo programs. 
        Predictions of a pQCD calculation \protect\cite{wang:1998:qcd} 
        are included as a solid line. The inset shows the ratios of the 
        results of the Monte Carlo codes to the experimental data.
		}
        \protect\label{fig:spectra}
\end{figure}

The measured neutral pion spectrum from  
central Pb+Pb reactions (10\% of min.bias cross section) as a function 
of $m_{T} - m_{0}$ is shown in Fig.~\ref{fig:spectra}. 
The data are compared to predictions of 
the string model Monte Carlo generators  
FRITIOF 7.02 \cite{ANDERSSON93} and VENUS 4.12 \protect\cite{WERNER93a}. 
As already observed in S+Au reactions \cite{wa80:pi0:98}, both 
generators fail to describe the data well at large $m_{T}$. The
FRITIOF prediction is more 
than an order of magnitude lower at high $m_{T}$ while VENUS 
significantly overpredicts the data. 
Alternatively,
it has recently been shown that perturbative QCD calculations, 
including initial state  
multiple scattering and intrinsic $p_{T}$ \cite{wang:1998:qcd}, are 
able to describe the preliminary WA98 data at intermediate and high 
$p_{T}$.  This prediction is included in Fig.~\ref{fig:spectra} as 
a solid line.\footnote{The results shown have been corrected for a 
small numerical error by the author of \protect\cite{wang:1998:qcd} 
and have changed by $\approx 10-30 \%$ compared to the publication.} 
The pQCD calculation  
shows a very good agreement in the high $m_{T}$ region. 
This surprising agreement has been interpreted as
an indication for unexpectedly small effects of parton energy loss
\cite{wang:1998:qcd}. On the other hand,  the 
parton cascade Monte Carlo code, VNI, which provides a more detailed
pQCD description, overpredicts the measured WA98 result by more
than a factor of ten at large $p_T$ \cite{GEIGER97}. 

An alternative picture of the ultra-relativistic heavy ion reaction
assumes that the mean-free path of the produced particles in the
overlap region is sufficiently small that a thermal or hydrodynamical 
interpretation may be applied. The excited matter is characterized
by thermal variables, such as temperature and  collective
velocities, which evolve over time. Hydrodynamical descriptions (see 
e.g. \cite{wiedemann96}) with properly adjusted parameters can 
describe the momentum spectra reasonably well.
This is, however, beyond the scope of this paper -- a detailed 
investigation of hydrodynamical descriptions of the momentum spectra 
will be given in a forthcoming publication. 

In view of the above discussion and the difficulty to describe the 
details of the neutral pion spectrum, it is apparent that the theoretical
description of ultra-relativistic nucleus-nucleus collisions remains
uncertain. In order to demonstrate a consistent description of
nuclear effects it is important to investigate the details of the pion
production as a function of the system size. 
To study the centrality dependence of the spectral shape
in a manner which is independent of model or fit function 
we have used the truncated mean transverse momentum 
$\langle p_T(p_{T}^{min}) \rangle$, where
\begin{equation}
        \langle p_T(p_{T}^{min}) \rangle = \left( \left.
        \int_{p_{T}^{min}}^{\infty} p_{T} \frac{dN}{dp_{T}} dp_{T} \right/
        \int_{p_{T}^{min}}^{\infty} \frac{dN}{dp_{T}} dp_{T} \right) -
        p_{T}^{min}.
        \label{eqn:mean-pt}
\end{equation}
The lower cutoff $p_{T}^{min} = 0.4$ GeV/$c$ is introduced to 
avoid systematic errors from extrapolation to low $p_{T}$ and 
has been chosen 
according to the lowest $p_{T}$ of the present data 
where systematic uncertainties imposed by the 
necessary corrections are still small. In general, the 
value of $\langle p_T(p_{T}^{min}) \rangle$ differs from the 
true average $p_{T}$, except in the case of a purely exponential 
distribution $d\sigma/dp_{T}$. 
For a purely exponential invariant cross section,
$d^{2}\sigma/dp^2_{T}$, $\langle p_T(p_{T}^{min}) \rangle$
decreases with increasing $p_{T}^{min}$.

\begin{figure}[tbh]
    \centerline{\includegraphics{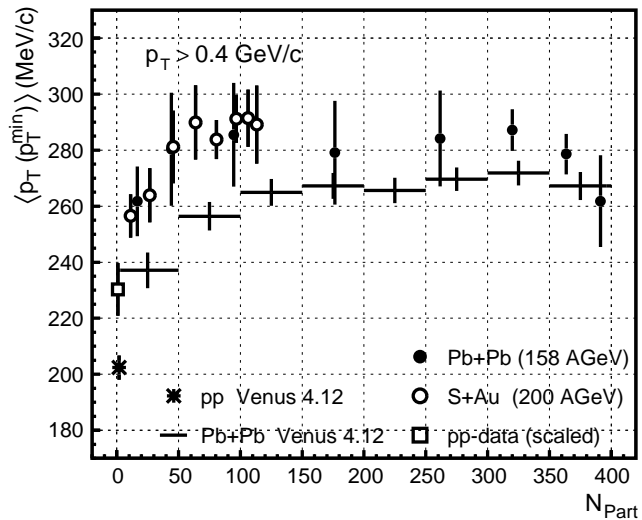}}
    \caption{Truncated mean transverse momentum 
    $\langle p_T(p_{T}^{min}) \rangle$ 
    of $\pi^0$ mesons as defined by 
    Eq.\,\ref{eqn:mean-pt} plotted as a function of the average number of 
    participants $N_{part}$. The solid circles 
    correspond to the 8 $E_{T}$ based centrality selections for Pb+Pb.
    The open square shows 
    $\langle p_T(p_{T}^{min}) \rangle$ 
    extracted from a parametrization of pp data scaled 
    to the same cms-energy \protect\cite{phd:blume::98}, the open 
    circles the results for S+Au collisions at 200 $A$GeV 
    \protect\cite{wa80:pi0:98}.  For comparison, results from 
	Venus 4.12 \protect\cite{WERNER93a} are included as 
    histograms for Pb+Pb collisions and as a star for pp. 
	A cut parameter $p_{T}^{min} = 0.4 \, {\mathrm{GeV}}/c$ was used.}
    \protect\label{fig:meanpt}
\end{figure}

Figure~\ref{fig:meanpt} shows $\langle p_T(p_{T}^{min}) \rangle$ as a 
function of the average number of participants $N_{part}$ for 158$\cdot$A~GeV 
$^{208}$Pb+Pb collisions. For comparison,  
$\langle p_T(p_{T}^{min}) \rangle$ values for 
200$\cdot$A~GeV  S+Au \cite{wa80:pi0:98} and from a parametrization of pp 
data \cite{phd:blume::98} are also included. 
$N_{part}$ is extracted by the assumption of a
monotonic relation between impact parameter and  
transverse energy and using the resulting correspondence between 
measured cross section and impact parameter. 
The average number of participants is calculated 
from nuclear geometry using the extracted impact parameter. 
Together these data show the 
general trend of a rapid increase of
$\langle p_T(p_{T}^{min}) \rangle$ compared to pp results for small 
system sizes. 
For $N_{part}$ greater than about 30 the mean transverse momentum 
appears to attain a limiting value of $\approx 280 \, {\mathrm{MeV}}/c$.
Venus 4.12 \cite{WERNER93a} calculations show a qualitatively  
similar behaviour, although the values of $\langle p_T(p_{T}^{min}) \rangle$
are somewhat lower than the experimental data.
The simple implementation of rescattering which is used in this model 
seems to be strong enough to lead to a saturation for semi-peripheral 
collisions as in the experimental data. One should, however, keep in 
mind that Venus 4.12 does not correctly describe pion production at 
high $p_{T}$ (see figure~\ref{fig:spectra}).

Alternatively, 
exponential fits to the spectra  over the range
$0.7 \, {\mathrm{GeV}}/c^2 \le m_{T} - m_{0} \le 1.9 \, 
{\mathrm{GeV}}/c^2$ have been performed. Although they do not 
adequately describe the spectra, even in this limited transverse mass 
range, they provide another measure of the spectral shape.  The 
fitted inverse slopes show a similar saturation effect  \cite{phd:blume::98}.

Earlier investigations of the dependence of $\langle p_T \rangle$ of
pions on system size 
\cite{wa80:pi0:98,wa80-gamma-87,helios-90} at SPS energies
have suggested such a saturation for large systems. 
The present study is the first investigation of the dependence
with Pb ions at the SPS.
Preliminary results from the AGS have indicated a weak increase
in the average $m_T$ of pions with the number of participants for 
Au+Au collisions \cite{e866}.

It is important to note that the observed limiting 
behaviour is very different from 
the observations in pp or $\mathrm{p \bar{p}}$ 
collisions. 
For very high energy $\langle p_T \rangle$
rises with the pseudorapidity 
density of charged particles 
\cite{sfm-87,ua1-82,cdf-88,e735-94}. 
In that case, more violent parton
scatterings presumably result in a harder spectrum of leading
particles together with a greater multiplicity of fragmentation
products. This would lead to the observed correlation between
$\langle p_T \rangle$ and multiplicity.
At lower $\sqrt{s}$, comparable to the data presented here, 
$\langle p_T \rangle$ decreases for increasing 
multiplicity \cite{kafka-77}, most likely due to energy 
conservation. In the case of nuclear
reactions, this anti-correlation 
is lost due to the large number of binary  
collisions. Instead, the initial increase of $\langle
p_T(p_{T}^{min}) \rangle$  with $N_{part}$ 
is interpreted as a result of multiple scattering. Initial state 
multiple scattering, as suggested as explanation for the Cronin
effect \cite{cronin}, would imply a continuing increase of $\langle
p_T(p_{T}^{min}) \rangle$ for more central collisions.  
Here however, the surprising observation is that additional
multiple scattering, implied by increasing $N_{part}$, does not alter the 
pion distributions. This is most easily understood as a 
consequence of final state rescattering and is, of course, 
the behaviour expected for a
thermalized system. 

\begin{figure}[bth]
	\centerline{\includegraphics{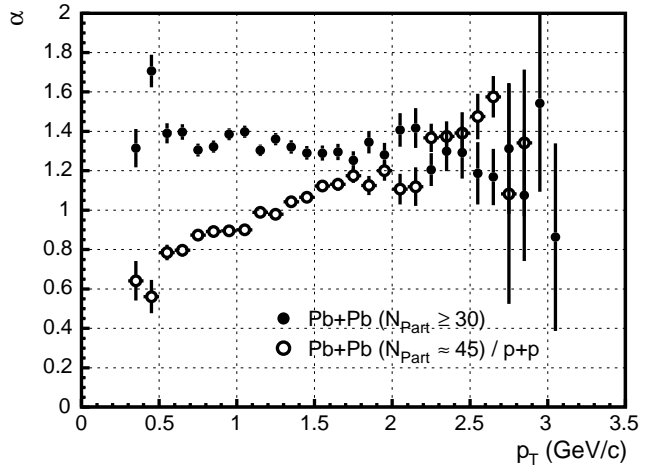}}
        \caption{    The exponent $\alpha(p_T)$ of the dependence
        of the $\pi^0$ yield on the average number of participants $N_{part}$
plotted as a function of the transverse
    momentum for 158 $A$GeV Pb+Pb. The solid circles 
    are calculated based on the centrality selections 
with $N_{part}\geq 30$. 
The open circles are calculated based on the ratio of the semi-peripheral 
data ($N_{part} \approx 45$) to a parameterization of pp data. 
}
        \protect\label{fig:alpha}
\end{figure}

More detailed information about the centrality dependence of the 
pion spectral shape and yield is shown in Fig.~\ref{fig:alpha} where
the neutral pion yield per event has been parameterized as
$E d^3N/dp^3 \propto N_{part}^{\alpha(p_T)} \cdot \sigma_0(p_T)$.
The results for $N_{part} > 30$ are well described by this scaling
with an exponent $\alpha(p_T) \approx 1.3$, independent of $p_T$.
Consistent with the previous discussion, the results indicate 
a constant spectral shape over the entire interval of
measurement from $0.5 < p_T < 3$ GeV/c. 
The observed $N_{part}^{4/3}$ scaling for symmetric systems implies
a scaling with the number of nucleon collisions, as confirmed by a
similar analysis.
However, this scaling does
not extrapolate from the pp results.
On the contrary, when comparing semi-peripheral Pb+Pb collisions with 
pp the exponent $\alpha$ varies over the entire $p_{T}$ interval, 
confirming the very different spectral shapes.

In summary, we have analyzed the centrality dependence of  high precision 
transverse momentum spectra of neutral pions from 158$A$GeV Pb+Pb
collisions. The neutral pion spectra are observed to show increasing
deviation from pp results with increasing centrality, indicating
the importance of multiple scattering effects. However,
for centralities with more than about 30 participating nucleons, 
the shape of the
transverse momentum spectrum becomes invariant over the interval  
\mbox{$0.5 < p_T < 3$ GeV/$c$}. In this interval the pion yield scales
like $N_{part}^{1.3}$, or like the number of nucleon collisions,
for this range of centralities. Since the amount
of rescattering increases with centrality, the invariance of
the spectral shape with respect to the number of rescatterings, 
most naturally suggests a dominantly thermal 
emission process. 
It will be important to determine whether cascade models which 
reproduce the observed invariant spectral shape will support the
interpretation as an ``effective'' thermalization due to significant
rescattering.

We wish to express our gratitude to the CERN accelerator division for
excellent performance of the SPS accelerator complex. We acknowledge with
appreciation the effort of all engineers, technicians and support staff who
have participated in the construction of the experiment. 

This work was supported jointly by 
the German BMBF and DFG, 
the U.S. DOE,
the Swedish NFR and FRN, 
the Dutch Stichting FOM, 
the Stiftung f{\"u}r Deutsch-Polnische Zusammenarbeit,
the Grant Agency of the Czech Republic under contract No. 202/95/0217,
the Department of Atomic Energy,
the Department of Science and Technology,
the Council of Scientific and Industrial Research and 
the University Grants 
Commission of the Government of India, 
the Indo-FRG Exchange Program,
the PPE division of CERN, 
the Swiss National Fund, 
the International Science Foundation under Contract N8Y000, 
the INTAS under Contract INTAS-93-2773, 
ORISE, 
Research-in-Aid for Scientific Research
(Specially Promoted Research \& International Scientific Research)
of the Ministry of Education, Science and Culture, 
the University of Tsukuba Special Research Projects, and
the JSPS Research Fellowships for Young Scientists.
ORNL is managed by Lockheed Martin Energy Research Corporation under
contract DE-AC05-96OR22464 with the U.S. Department of Energy.
The MIT group has been supported by the US Dept. of Energy under the
cooperative agreement DE-FC02-94ER40818.


\end{document}